\documentclass[american,preprint, superscriptaddress]{revtex4}
\usepackage[T1]{fontenc}
\usepackage[latin9]{inputenc}
\usepackage{textcomp}
\usepackage{graphicx}
\usepackage{babel}

\begin{document}

\title{Protein micro patterned lattices to probe a fundamental lengthscale
involved in cell adhesion.}

\author{Hervé Guillou }

\affiliation{Institut Néel, CNRS UPR2940 et Université Joseph Fourier, BP166,
38042 Grenoble, France}

\author{Benoit Vianay}

\affiliation{Institut Néel, CNRS UPR2940 et Université Joseph Fourier, BP166,
38042 Grenoble, France}

\author{Jacques Chaussy}

\affiliation{Institut Néel, CNRS UPR2940 et Université Joseph Fourier, BP166,
38042 Grenoble, France}

\author{Marc R. Block}

\affiliation{INSERM U823, Equipe DySAD, Institut Albert Bonniot, Site Santé, BP170,
38047 LaTronche, France}

\thanks{author's fax: 33+456 38 70 87}
\email{herve.guillou@grenoble.cnrs.fr }

\begin{abstract}
Cell adhesion, a fundamental process of cell biology is involved in
the embryo development and in numerous pathologies especially those
related to cancers. We constrained cells to adhere on extracellular
matrix proteins patterned in a micro lattices. The actin cytoskeleton
is particularly sensitive to this constraint and reproducibly self
organizes in simple geometrical patterns. Such highly organized cells
are functional and proliferate. We performed statistical analysis
of spread cells morphologies and discuss the existence of a fundamental
lengthscale associated with active processes required for spreading.
\end{abstract}

\maketitle
In vivo, cells proliferate on the 3D structure of extracellular proteins
that forms a highly organized scaffold: the extracellular matrix (ECM)
is an arrangement of different proteins\cite{alberts}.
Cell adhesion onto the ECM is a fundamental process involved in the
embryo development as well as in numerous pathologies especially those
related to cancers \cite{thiery2003}. In many cell types, the integrin
family of transmembrane proteins is responsible for cell interactions
with ECM proteins. This specific interaction among
other micro environmental physicochemical parameters such as support
compliance \cite{discher2005} or shape \cite{huang1999,Chen1997,thery2006}
has been shown to be relevant for the regulation of cell proliferation
and to the accomplishment of their functions.

Following \cite{huang1999,Chen1997,thery2006} we use adequately tailored micro lattice of proteins to obtained very
reproducible and simple actin cytoskeleton organization.  The microfabrication of the
protein array and the simplified actin cytoskeleton morphology that
we systematically observe are described next. The result of the statistical analysis of cell shapes
are presented in a second part. In particular, we discuss results indicating the existence
of a characteristic length. This length is associated with the actin
based, active protrusive processes needed to explore the environment
and to establish new adhesive contacts required for spreading and
migration.

\begin{figure}
\begin{centering}
\includegraphics[width=1\columnwidth]{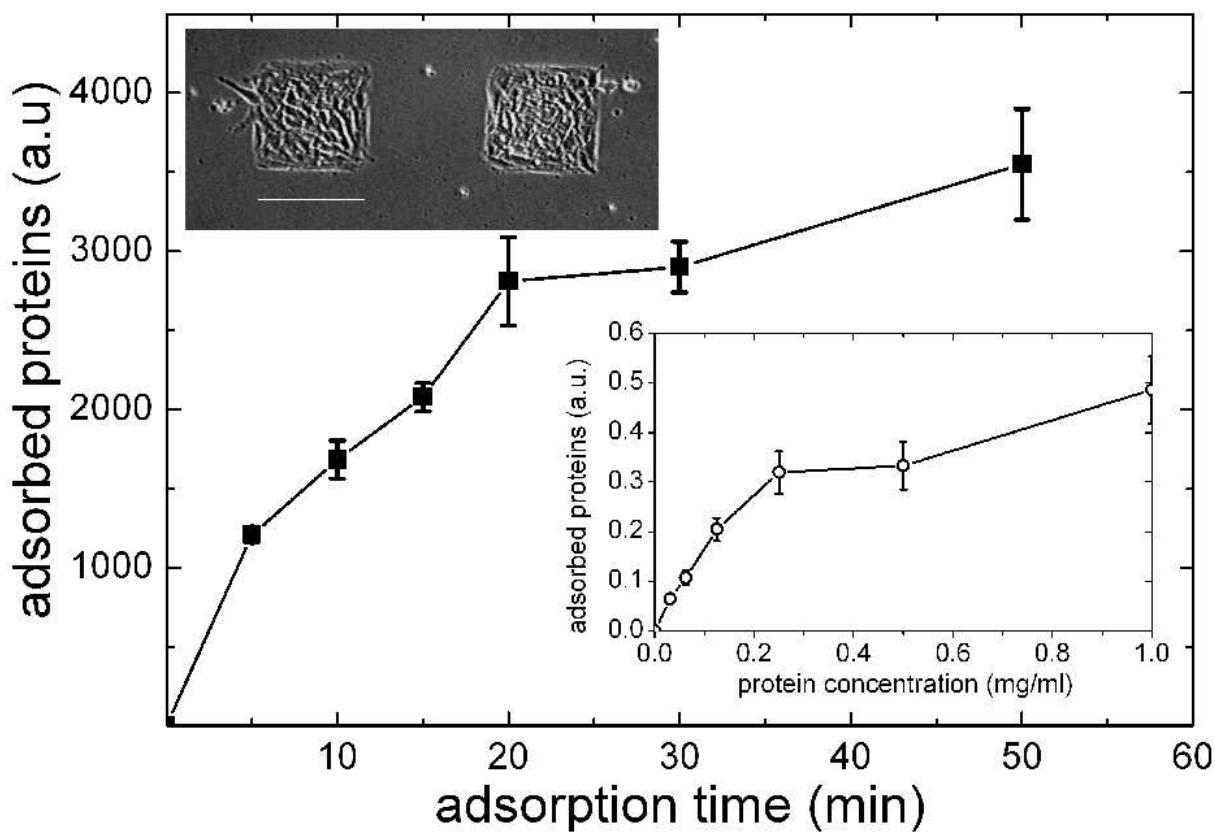}
\par\end{centering}
\caption{\label{fig:guillou-fig1}Kinetic of fibronection adsorption on hydrophobic
glass substrates. Lower inset: adsorption isotherm on hydrophobic
surface. Upper inset: durability of protein patterns with cells cultured
on fibronectin square islands.}
\end{figure}
According to amphiphilic properties of globular proteins they tend
to accumulate and adsorb at interfaces \cite{israelachvili1991}. Hydrophobic
surfaces yield protein films of larger density than hydrophilic surfaces
\cite{sorribas2002}. The
principle of fabrication of the protein lattice is based on such non
specific protein immobilization on hydrophobic surfaces through a
patterned photo-resist mask followed by a lift-off. The photo-resist
is insolated using optical UV lithography. Patterns with a critical
size of $1\mu\mbox{m}$ are thus achievable. In this study, although
several proteins where successfully tested (fibronectin, fragment
of fibronectin, antibodies, vitronectin, gelatin) best results in
terms of spatial resolution were obtained with a fragment of fibronectin
made of domains of type III 7 to 10\cite{Leahy1996}.
Fig. \ref{fig:guillou-fig1} shows the adsorption kinetic of the fragment
at $20\mu\mbox{g}/\mbox{ml}$. The adsorption isotherm on hydrophobic
surface is shown in the inset of fig. \ref{fig:guillou-fig1}.  Both graphs
indicate that the adsorption is limited as expected \cite{guemouri2000}. After adsorption,
the photo-resist mask is lifted-off.
The yet photo-resist protected surface must be blocked to prevent
further adsorption of protein from the culture medium that would otherwise
rapidly blur out the protein pattern. This major problem
in protein patterning and has been solved \cite{zhang1998,liu2002}
by using a commercially available polymer \cite{pluronic.} that
spontaneously adsorb on hydrophobic surface in the same manner as
proteins. We were able to maintain consistent cell patterns and to
keep the repellent properties of the surface for more than 4 days
as shown in the images of figure \ref{fig:guillou-fig1}: the square
adhesive islands kept their geometry despite confluent cell crowding.
The lift-off of proteins on hydrophobic surfaces followed by the immobilization
of blocking polymer is a very efficient way to produce high resolution
adhesive patterns of controlled geometry. Nevertherless it will never
reach the resolution achieved in \cite{Cavalcanti-Adam2007} where
patterning at the molecular level is realised. However, in contrast
with microstamping it produces patterns of very high quality and allows
easy alignement with existing structures. 

\begin{figure}[h]
\begin{centering}
\includegraphics[width=1\columnwidth]{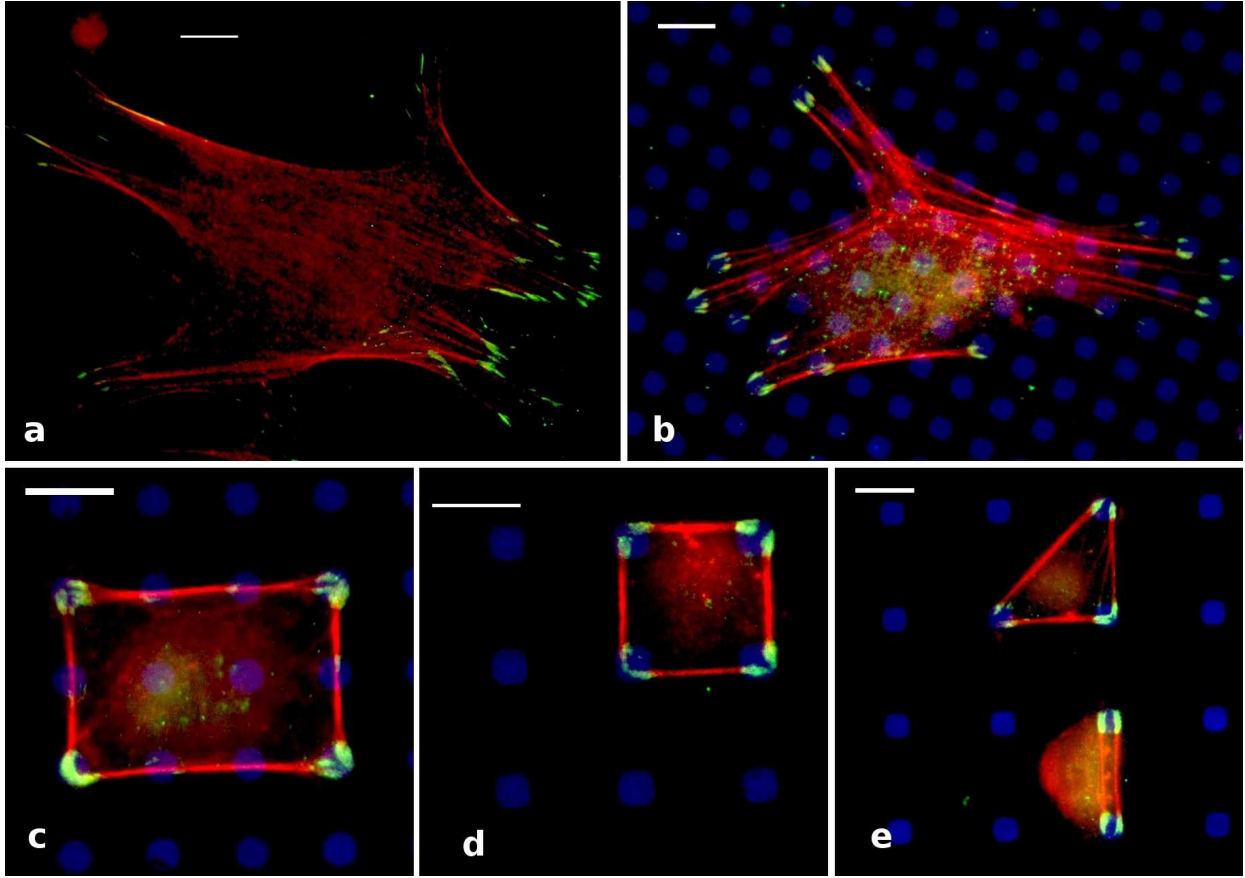}
\par\end{centering}
\caption{\label{fig:different-shapes}Typical single cell shapes observed for
various protein lattice. Red is F-actin cytosqueleton, green is vinculin
and blue is fibronectin. Scale : white bar is $10\mu m$. Non patterned
substrate (a), lattice pitch $4\mu m$ (b), $6\mu m$ (c), $10\mu m$(d),
$14\mu m$ (e). }
\end{figure}
\noindent We tested the biocompatibility of the protein lattice by
studying cell spreadings. Cells were harvested and plated onto the
patterns and spread for 4 hours before fixation in paraformaldehyde.
The actin cytoskeleton was revealed using rhodamine labelled phalloidin
and focal contacts by primary antibody directed toward vinculin. Cover
slips were mounted in mowiol for fluorescence microscopy. More details
on the methods are given in \cite{Guillou2008}. For non patterned substrates
(fig. \ref{fig:different-shapes}a) no particular actin cytoskeleton
organization can be observed. For small distances between adhesive
dots (fig. \ref{fig:different-shapes}b), the cytoskeleton organisation
is similar to the one observed on non patterned substrates. However,
as the distance between adhesive spots is increased, the shape of
cells is very quickly simplified into elementary geometries (fig.
\ref{fig:different-shapes}c-d). It is remarquable how the
complex cytoskeleton organization of fig.s \ref{fig:different-shapes}a-b
is simplified: stress fibers are exclusively located on the edges
of the cell body and linked to well identified focal contacts at their
ends. 

\begin{figure}[h]
\begin{centering}
\includegraphics[width=1\columnwidth]{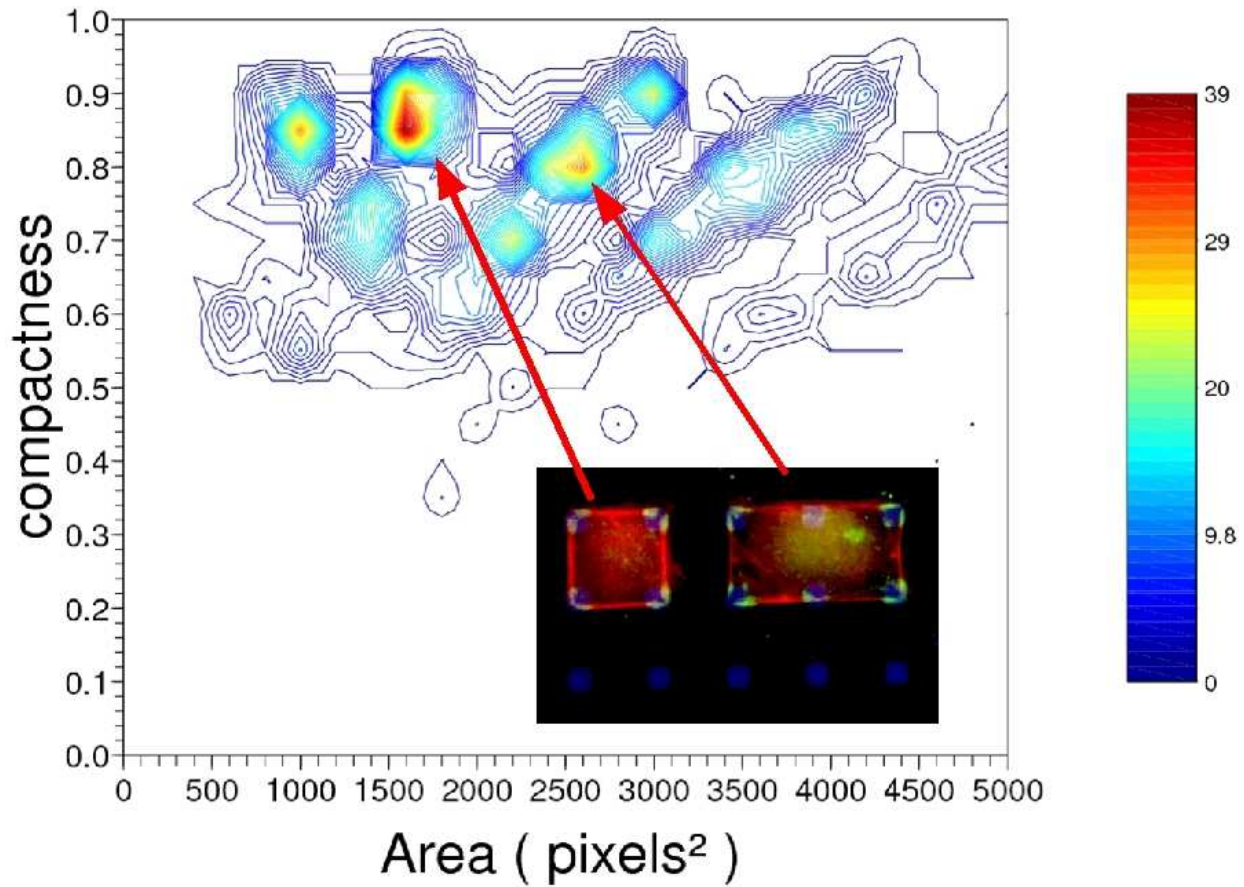}
\par\end{centering}
\caption{\label{fig:shape-classification}Distribution of shape obtained for
a dot spacing of $10\mu m$. The colorcode represent the number of
cells observed in a particular configuration. The maxima of occurence
are obtained for the 4 dots square and 6 dots rectangle shown in the
inset.}
\end{figure}

Each actin cytoskeleton
organization was classified according to its area
and compacity \cite{compacity}. Fig. \ref{fig:shape-classification}
shows the population of cells in each class for a lattice
pitch of 10 \textmu{}m. 
Each peak corresponds
to a simple geometrical compact shape that can be drawn between adhesive
islands.  It shows that the most probable organization is the
4 dots square and that cell shapes can be classified.

\begin{figure}[h]
\includegraphics[width=1\columnwidth]{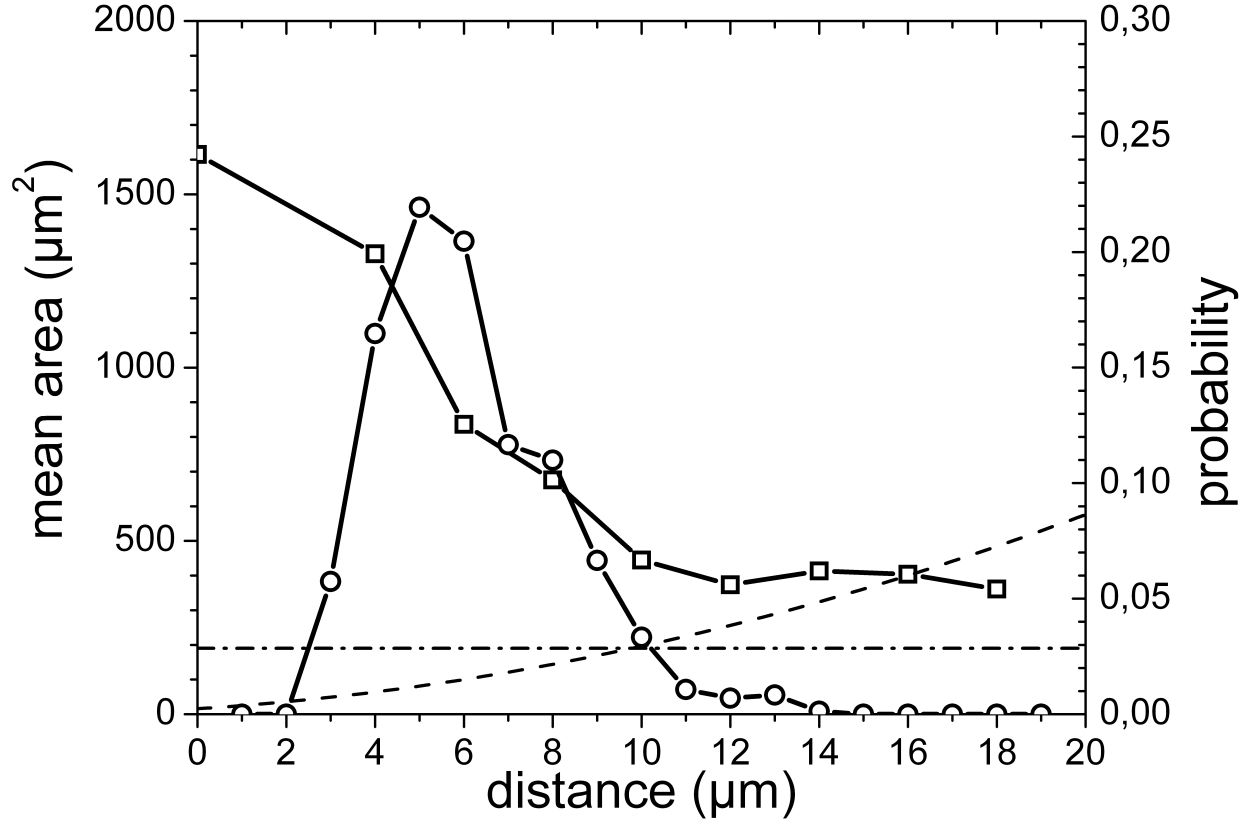}
\caption{\label{fig:filopodia}Shape selection through protrusion length. Left
axis, open squares: average mean area of single spread cells on proteins
arrays with different pitches. Right axis, open circles: probability
to find a filopodia of a given length (averaged on 20 different cells
and 400 filopodia.}
\end{figure}
On a non patterned substrate the
ECM layer is dense and almost every membrane protrusion leads to engagement
of integrins. Very quickly cells adopt a migrating phenotype in which
the cytoskeleton is continuously rearranging. Cell shapes cannot be classified.
 By contrast, on a protein lattice, a finite
distance between neighbouring adhesive spots is introduced and the
membrane protrusion are less efficient in finding new ligands. Fig.
\ref{fig:filopodia} supports this idea and compares the mean area of spread cells as a function
of the lattice pitch with the distribution of the length of filopodia.
The mean area decreases very rapidly once the distance is increased.  As shown
on fig. \ref{fig:filopodia} membrane protrusions are able to explore
only the close environment of cells. The probability that a protrusion
exceeds a distance of $10\mu\mbox{m}$ is low. Moreover, because of
the geometry of the lattice, all protrusions are not successful in
finding new ligand, reducing further the probability to engage membrane
receptors. The distance between adhesive spots introduce thus a length
that can be compared with the average length of protrusion. %
By increasing the distance between adhesive spots the probability
to change the set of adhesive constraint is gradually reduced. The
actin cytoskeleton has thus time to relax and to self organize into
very simple shapes. The protein lattice probe thus a fundamental length
related to the protrusion length.

In this letter we studied single cell spreading onto micro fabricated
patterns of controlled geometries. Protrusions 
are the limiting
process resulting in a greatly reduced spread cell area as soon as
the pitch is greater than their typical length. As a result stress fibers are
exclusively located at the edges of the cell body and attached to
two focal adhesion at their ends. 

\bibliographystyle{unsrt}

\begin{thebibliography}{10}
\bibitem{compacity}
compacity is defined as in the IMAQ library of the Labview programming software
  from National Instrument, Austin, Texas, USA.
\end{thebibliography}

\end{document}